\documentclass[
11pt,%
tightenlines,%
twoside,%
twocolumn,%
nofloats,%
nobibnotes,%
nofootinbib,%
superscriptaddress,%
noshowpacs,%
centertags]%
{revtex4-2}

\usepackage{amssymb}
\usepackage{amsfonts}
\usepackage[tbtags]{amsmath}
\usepackage{amscd}
\usepackage{amscd}
\usepackage{amsthm}
\usepackage{mathtext}
\usepackage[T1,T2A]{fontenc}
\usepackage[utf8]{inputenc}
\usepackage[english,russian]{babel}
\PassOptionsToPackage{pdfauthor={Vladimir V. Kisil},%
    pdftitle={Interpretation of wave function by coherent ensembles of trajectories},%
    pdfsubject={physics},%
    pdfkeywords={quantum mechanics, wave mechanics. particles}, %
    breaklinks=true,%
      colorlinks=true,%
      bookmarks=true,%
}{hyperref}
  \AtBeginDocument{\hypersetup{breaklinks=true,%
      colorlinks=true,%
      bookmarks=true,%
      pagebackref=true
}}
\usepackage{hyperref}
\usepackage{amsthm}

\providecommand{\rmi}{\mathrm{i}}
\providecommand{\rme}{\mathrm{e}}
\providecommand{\rmd}{\mathrm{d}}

\theoremstyle{definition}

\newtheorem{definitionR}{Определение}
\newtheorem{definition}{Definition}

\hypersetup{breaklinks=true
}
\begin{document}

\selectlanguage{english}

\title{Interpretation of wave function by coherent ensembles of  trajectories}

 \author{\firstname{Vladimir V.}~\surname{\href{http://v-v-kisil.scienceontheweb.net/}{Kisil}}}
 \email[E-mail: ]{{V.Kisil@leeds.ac.uk}}
 \thanks{Home page: \url{http://v-v-kisil.scienceontheweb.net/}. On leave from the Odessa University. \\
   Paper is accepted in \href{https://link.springer.com/journal/11497}{\emph{Physics of Particles and Nuclei Letters}}, ©2025 \url{http://pleiades.online/}.}

 \affiliation{School of Mathematics,
 University of Leeds,
 Leeds LS2\,9JT,
 England}


\begin{abstract} 
  We re-use some original ideas of de~Broglie, Schr\"odiger, Dirac and Feynman to revise the ensemble interpretation of  wave function in quantum mechanics. To this end we introduce coherence (auto-concordance) of ensembles of quantum trajectories in the space-time. The coherence condition accounts phases proportional to classical action, which are in foundation of the Feynman path integral technique. Therefore, our interpretation is entirely based on  well-known and  tested concepts and methods of wave mechanics. Similarly to other ensemble interpretations our approach allows us to avoid all problems and paradoxes related to wave function collapse during a measurement process. Another consequence is that no quantum computation or quantum cryptography method will ever work if it assumes that a particular q-bit represents the entire wave function.     
\end{abstract}

\maketitle

\hfill\parbox[t]{.4\textwidth}{\small 
\emph{History of science is not a cemetery with tombs over ideas buried forever, instead it is a collection of unfinished architectural ensembles, many of them were not completed not for deficiencies of the project but because of a lack of technical or economical demands.}\\
\centerline{A.A. Lyubischev}}\medskip

\section{Introduction}
\label{sec:introduction-2}

Recently we passed centennial of few short papers of Louis de~Broglie, which invented \emph{wave mechanics}---also known as \emph{quantum} theory.  This opposition of names reflects the \emph{wave--particle duality}  proposed by de~Broglie---a fundamental property generally accepted since then. The discovery of wave properties of matter opens the door for unprecedentedly rapid development of theory, in particular paved the way for the Schr\"odinger \emph{wave equation}~\cite{Schrodinger26b}. However, rereading de~Broglie's papers we may spot some ideas which were abandoned in the past even by their author himself. However, those guesses deserves to be revised and may be even rehabilitated.

Mathematical foundations of quantum mechanics---the theory of operators in Hilbert spaces---were extensively studied and firmly placed in physical research. Yet, there are continuing discussion of  interpretations of the wave function and related issues like a \emph{wave function  collapse} during a measurement process. Corresponding questions remain a topic of on-going (although not so central now) debates. The discontent provokes some radical constructions like the many-world interpretation~\cite{Everett57}. Since such remedies are not better than the disease itself, many scientists choose  to marginalise interpretation questions under the slogan ``shut up and calculate!'~\cite{Mermin04a}'.   For example, textbook~\cite{GriffithsSchroeter18a} dedicates its main 11 chapters to mathematical content of quantum theory and only the final Ch.~12 considers measurements, (non-)locality, causality, Einstein--Podolsky--Rosen paradox, Schr\"odinger cat and other controversial topics.

One of the route to avoid issues created by the wave function collapse is an \emph{ensemble interpretation}, which is often attributed to Einstein~\cite{Khrennikov12a} or other physicists~\cite{Pechenkin04a,Pechenkin22a}. However, let us read the final paragraph  of de~Broglie's 1922 paper~\cite{deBroglie22b}:
\begin{quote}
  \small Au point de vue des quanta de lumière, les phénomènes d’interférences paraissent liés à l'existence d’agelomérations d'atomes de lumière dont les mouvements ne sont pas indépendants, sont cohérents. Dès lors, il est naturel de supooser que si la théorie des quanta de lumière parvient un jour à interpréter les interférences, elle devra faire intervenir de telles agglomérations de quanta.
\end{quote}
Clearly, de~Broglie connects wave phenomena of matter with a group behaviour of coherent ensemble of particles few years before the wave function was invented.
This ensemble (aka \emph{statistical}) interpretation was often reinvented and revised, cf.~\cite{Blokhintsev51a,Ballentine70a} and already cited surveys~\cite{Khrennikov12a,Pechenkin04a,Pechenkin22a}.  However, the presented arguments often leave a space for relentless attacks from the  position of Copenhagen dogma, in particular, by such notable physicists as V.A.~Fock~\cite{Fock37a,Fock52a,Fock57a}. The weakest place was the definition of quantum ensemble itself and its property to be coherent. We are addressing these topics below.

\section{Quantum ensambles}
\label{sec:quantum-ensambles}

We start from clarifications based from earlier guesses of founding fathers:
\begin{enumerate}
\item Initially in~\cite{deBroglie23} de~Broglie deduced wave properties of quanta from the assumption that each of them possesses some \emph{inner periodic process}. A similar theme with individual ``stopwatch'' was used by Feynman in the popular lectures on path integrals~\cite{Feynman88qedRus}. A careful consideration~\cite{Kisil12cRus,Kisil17a} shows that it is the periodicity of the function \(\rme^{\rmi t}\)  produces quantum mechanics---regardless of the Dirac postulate on ``noncommutativity of quantum observables''~\cite{Dirac26b}.
\item The same de~Broglie's paper~\cite{deBroglie23} introduces a \emph{fictional} spatial wave such that it phase at every point coincides with the phase of the above inner periodic process of a quantum located there. This assumption on \emph{stability} (saying differently \emph{coherence}) allowed de~Broglie to deduce Bohr's quantisation rules for the first time.
\item \label{item:whole-coherence}
  From the above coherence condition between the phase of the fictional wave and the inner periodic process we can conclude that any two coherent quanta arrived to the same point of space-time shall share the same phase. Similarly, any two quanta departing from a same point in different directions start their voyages with a common phase.
\item De~Broglie's ideas were used by Schr\"odinger in~\cite{Schrodinger26b} to justify his wave equation using the Hamilton optic-mechanical analogy~\cite{deBroglie26a} between ray optics and classical mechanics\footnote{Having the excellent textbook~\cite[\S\,46]{Arnold89Rus} it is hard to believe today that Schr\"odinger called that analogy ``nearly forgotten''. However, we shall not be too much surprised because the Schr\"odinger own justification of the wave equation is now missing from textbooks as well: the equation is either simply declared or refereed to the similarly declared ``quantisation rules''.}: Schr\"odinger assumed that quantum theory is similar to more physical undulatory  optics.  This correspondence suggests that the frequency of inner periodic process is changing over the time and its time derivative is proportional to the Lagrangian function of the particle.
\item  If the frequency is proportional to the Lagrangian \(L\) then the change of the phase over a path \(\gamma\) shall be proportional to the action function \(S=\int_\gamma L\,\rmd t\). It was explicitly written by Dirac in~\cite{Dirac33a} and later becomes the foundation of the Feynman path integral method~\cite{Feynman88qedRus,FeynHibbs65}. Notably, Feynman's approach initially got a cool reception from Copenhagen adepts for the usage of tabooed concept of ``trajectory''. The denial was softened later under the agreement that a single trajectory is only an imaginary computational device and ``in reality'' every quantum is passing over all  paths simultaneously. Obviously, the authoritative textbook~\cite{FeynHibbs65} may not say that the phase \(S/\hbar\)  is assigned to some inner periodic process of a particle travelling along its individual path. However, exactly such image is read from ``we have a stopwatch that can time a photon as it moves''~\cite{Feynman88qedRus} while the author is less locked within the official dogma.
\end{enumerate}
The coherence condition in item~\ref{item:whole-coherence} may be too restricitve. In practice the exact match of phases may be relaxed till their \emph{sufficient} closedness. Correspondingly, one may define a measure of a decoherence of an ensemble adding up all phase difference in the intersection points of different trajectories. It may be possible to introduce an interaction between collating quanta which leads to increasing coherence of the entire ensemble, similarly to a physical body dynamics  towards its thermostatic equilibrium.

Therefore, we are arriving to:
\begin{definition}
  A \emph{quantum ensemble} consists of particles' trajectories in a certain region of  space-time. To every point of a trajectory we assign a \emph{phase} proportional to the physical action. In a sufficiently \emph{coherent ensemble} phases of all trajectories passing a particular point of space-time are about the same.
\end{definition}
Our discussion in terms of particles' trajectories rather than just their position at a specific moment of time is similar to ensembles of \emph{quantum processes} of K.V.~Nikolsky~\cite[\S3, \S12]{NikolskyKV41}. 

It is well-known that the Feynman path integral  produces the wave function of a particle coinciding with one obtained from the Schr\"odinger equation or the Heisenberg--Dirac quantisation~\cite[\S\,3.4]{FeynHibbs65}. Furthermore, the wast majority of trajectories do not contribute towards the wave function because they are arriving to a point with almost opposite phases. Thus, we are obtaining the same wave function if all such mutually cancelling paths will be dropped from our consideration. The remaining paths shall have phases which are close to the resulting wave function, thus those paths will form a sufficient coherent ensemble. Thereafter, we conclude that \emph{for every wave function there is a sufficiently coherent ensemble which produces this wave function by Feynman integration}.

It is common to object the ensemble interpretation by consideration of a \emph{very sparse ensemble}. For example, considering two slits interference one observes electrons travelling ``one-by-one'': the source does not emit a new electron until the previous one will not be absorbed by the detector~\cite{AspectGrangier87a}. Indeed, there is no path intersections in this case and there is no necessity (and possibility) to coordinate phases of electrons. However, in  such an experiment we also have the device causing interference. It interacts with all participating electrons and serves as a media for phase coordination. A principal possibility to model wave-type effects for such sparse ensembles was demonstrated in~\cite{Kisil01c,KhrenVol01,KisilHodgson21a}.

\section{Discussion and applications}
\label{sec:disc-appl}
Wave-particle duality of the matter remains a disputed topic in quantum theory. Among various treatments of the issue there exist some rather radical, e.g. to cancel corpuscular properties of light from the physical reality and shelve the notion of \emph{photon} next to obsolete XIX century \emph{ether}~\cite{Lamb95a}.

In this paper we revise the ensemble interpretation of the wave function with some original suggestions of de~Broglie reinforcing them with the later technique from the Feynman path integral.
Here are few immediate consequences of the proposed interpretation:
\begin{enumerate}
\item An observation of wave-type behaviour among quantum (discrete) object has statistical nature and, therefore,  is group property of coherent ensembles. Wave effects cannot be demonstrated by a single particle or by a group of completely disconnected particles. Similarly, the velocity of a single gas molecule does not tell the whole heat distribution in the considered volume, although the molecule makes its contribution to the picture. This analogy may turn to be not so superficial if we consider the Schr\"odinger equation as the heat equation with complexified time. 
\item The wave function corresponds to an ensemble of coherent quanta: its magnitude depends on the spatial-temporal  density of localised particles. The phase of the wave function at a particular point of the extended configurational space indicates the phase  of any quantum localised there and effectively serves as the de~Broglie's ``fictional wave''~\cite{deBroglie23}.
\item An observation/measurement act on a single particle returns its specific physical parameters, e.g. spatial-temporal localisation. This may cause a small perturbation of the containing ensemble. However, there is no a wave function collapse as a result of a single measurement since the wave function describes the whole coherent ensemble~\cite[\S14]{Blokhintsev83a}.
\item Therefore, the interpretation removes a ground for Einstein--Podolsky--Rosen paradox, the Schr\"odinger cat's dilemma, etc.~\cite[p.364]{Mandelshtam72}.
\item There is some physical reality behind phases (e.g. through some inner periodic processes) of individual quanta travelling along definite paths. In a contrast, the wave function is only integral quantity of all such individual phases within a coherent ensemble. Thereafter, there is no need to search for a physical media which conveys the wave function, cf. the analogy with the concept of heat distribution within a physical body mentioned above.
\item Similarly, we do not need any additional ``pilot waves'' which will allow a particle to have a trajectory agreed with the wave function. ``Double solutions'' of this type~\cite{deBroglie70a} were considered by de~Broglie once Copenhagen folks forced him to abandon his initial approach~\cite[Ch.\,6]{Lochak99a}.    
\end{enumerate}

Clearly, our discussion uses only traditional tools of quantum theory: the Schr\"odinger wave equation and Feynman path integrals ignited by some earlier guesses of de~Broglie. Therefore, there is no need for a separate detailed verification that our interpretation does agree with the established mathematical methods of wave mechanics.

Notably, the present consideration is similar to A.~Khrennikov's approach to the ensemble interpretation~\cite{Khrennikov12a}: first a certain causal and deterministic pre-quantum model is invented and then stochastic quantum picture is obtained by a reduction to a wave function. However, we do not (in contrast to~\cite{Khrennikov12a}) introduce a new pre-quantum theory; instead, we recombine well-known elements already present in place.

Despite of our discussion being on ``just the interpretation question'', some very practical conclusions may be deduced. Quantum computing and quantum cryptography are actively researched now, often the proposals assume that a single q-bit can hold the entire wave function of a quantum state.  These studies are generously funded for a substantial period of time due to prospective financial and military applications. Therefore, the absence of any practically working  devices of this type may be considered as \emph{experimental evidences} of various theoretical objections expressed earlier~\cite{Kisil09b,Dyakonov12a,Dyakonov18a,KisilHodgson21a}. In particular, such technology will not be ever implemented if the correct interpretation of the wave function is based on the group behaviour of a coherent ensemble. In other words, the failure to build a much anticipated quantum computer is a confirmation of the ensemble interpretation just like we learnt the energy conservation through all (often very clever) attempts to invent a perpetuum mobile. 

Finally, we observe from the history of de~Broglie's ideas that the list of reasons of incomplete scientific developments in the paper's epigraph shall be extended: often projects are abandoned for subjective reasons only.
\begin{acknowledgments}
  The author is grateful to the anonymous referee for useful suggestions which helped to improve the text. Prof.~Pechenkin directed our attention to several relevant sources.
\end{acknowledgments}

\clearpage

\selectlanguage{russian}

\onecolumngrid

\centerline{}
\begin{center}
  {\large
  \textbf{Интерпретация волновой функции и согласованные ансамбли траекторий}}\\[1em]
\firstname{В.В.}~\surname{\href{http://v-v-kisil.scienceontheweb.net/}{Кисиль}}
  \footnote{Емайл: \href{mailto:V.Kisil@leeds.ac.uk}{V.Kisil@leeds.ac.uk}\\
    Домашняя страничка: \url{http://v-v-kisil.scienceontheweb.net/}. В отъезде из Одесского университета.\\
  Статья принята для
публикации в \href{https://pepan.jinr.ru/index.php/PepanLetters}{<<Письма в журнал Физика элементарных частиц и атомного ядра. Теория>>}, ©2025 \url{http://pleiades.online/}.}\\[1em]
  \emph{Отделение математики, Университет г. Лидса, Англия}\\[1em]
\parbox[t]{.9\textwidth}{\small
  Мы используем некоторые изначальные идеи де~Бройля, Шрёдингера, Дирака и Фейнмана для уточнения ансамблевой интерпретации волновой функции в квантовой механике. Для этого рассматриваются согласованные ансамбли квантовых траекторий в пространстве-{}времени. Условие согласования основано на фазах пропорциональных классическому действию, знакомым всем из интеграла по траекториям Фейнмана. Таким образом, предложенная интерпретация в существенном использует только проверенные и хорошо зарекомендовавшие себя понятия и методы волновой механики. Как и другие ансамблевые интерпретации данный подход позволяет избежать всех парадоксов связанных с  редукцией волновой функции в процессе измерений. Ещё одно следствие, упоминаемое нами, это ожидаемая бесплодность всех попыток произвести квантовые вычисления или квантовое шифрование на методах предполагающих, что состояние отдельного кубита описывается полной волновой функцией. }
\end{center}

\twocolumngrid

\hfill\parbox[t]{.4\textwidth}{\small 
\emph{Прошлое науки---не кладбище с надгробными плитами над навеки похороненными идеями, а собрание недостроенных архитектурных ансамблей, многие из которых не были закончены не из-{}за несовершенства замысла, а из-{}за технической или экономической невостребованности.}\\
\centerline{А.А. Любищев}}\medskip

 \selectlanguage{russian}
\setcounter{section}{0}
\section{Введение}
\label{sec:introduction}
Недавно прошло столетие с  публикации нескольких коротких работ де Бройля, которые открыли эру \emph{волновой механики}---одного из наименований \emph{квантовой} теории.  Эта двойственность названий отражает предложенный де Бройлем  корпускулярно-{}волновой дуализм материи---фундаментальное свойство прочно утвердившееся с тех пор. Декларация волновых свойств материи было важным рубежом и послужило дальнейшему беспрецедентно быстрому развитию теории, в частности, стимулировало появление  \emph{волнового} уравнения Шредингера~\cite{Schrodinger26b}. Однако, перечитывая работы де Бройля по юбилейному поводу можно заметить некоторые идеи, которые были оставлены в прошлом даже самим их автором.  Тем не менее, эти догадки заслуживают быть вновь рассмотренными и, возможно, реабилитированными.

Фундаментальный математический аппарат квантовой механики---теория операторов в гильбертовых пространствах---хорошо разработан математически и занял прочное  место в физических исследованиях. Однако, остаётся очень спорный вопрос интерпретации волновой функции и связанных с этим различных допущений вроде \emph{редукции волновой функции}  в процессе измерений. Сопутствующие этому сложности продолжают быть предметом незатихающих (хотя и немного периферийных теперь) обсуждений. Эта неустроенность временами порождает весьма экзотические построения вроде многомировой модели~\cite{Everett57}. Так как такое лечение хуже самой болезни, то ответная реакция низводит вопросы интерпретации до второстепенных, дескать <<не болтай, а вычисляй!>> (``shut up and calculate!''~\cite{Mermin04a}). Например, в учебнике~\cite{GriffithsSchroeter18a} основные 11 глав посвящены математическому аппарату квантовой механики и только заключительная двенадцатая глава касается вопросов связанных с измерениями, локальностью, причинностью, парадоксом Эйнштейна--Подольского--Розена, котом Шрёдингера и другими весьма неоднозначными темами.

Одним из способов избежать проблем редукции волновой функции является её \emph{ансамблевая интерпретация}, которую часто отсылают к  Эйнштейну~\cite{Khrennikov12a} или другим физикам, см.~\cite{Pechenkin04a,Pechenkin22a}. Обратимся, однако, к  заключительному пассажу статьи де~Бройля 1922 года~\cite{deBroglie22b,deBroglie10r}:
\begin{quotation} \small
   С точки зрения квантов света явления интерференции кажутся связанными с существованием совокупностей атомов света, движущихся не независимо, а когерентно. Отсюда естественно предположить, что если когда-{}нибудь теория квантов света сможет объяснить интерференцию, она должна будет ввести в рассмотрение подобные объединения квантов.
\end{quotation}
Легко заметить, что де~Бройль связывает волновые свойства материи с групповым поведением согласованного (когерентного) ансамбля частиц ещё за несколько лет до появления волновой функции. Такая  {ансамблевая} (или \emph{статистическая}) интерпретация в дальнейшем  многократно переоткрывалось  и переосмысливалось многими авторами, см.~\cite{Blokhintsev51a,Ballentine70a} и ранее упомянутые обзоры~\cite{Khrennikov12a,Pechenkin04a,Pechenkin22a}.
Однако, предлагаемые формулировки 
зачастую оставляли пространство для неустанной критики с позиции копенгагенской интерпретации квантовой механики, в частности, и такими видными физиками как В.А.~Фок~\cite{Fock37a,Fock52a,Fock57a}. Наиболее уязвимым звеном оставалось само понятие квантового ансамбля и то, каким образом определялась его согласованность. Этот момент мы и рассмотрим ниже.

\section{Квантовые ансамбли}
\label{sec:-quantum-ansambles}

Начнём прояснять имеющиеся затруднения используя некоторые наблюдения из работ первооткрывателей:
\begin{enumerate}
\item Изначально в~\cite{deBroglie23} де~Бройль выводил волновые свойства квантов из наличия у каждого из них \emph{внутреннего периодического процесса}. Этот же мотив с индивидуальными <<часами>> использовался Фейнманом в популярном изложении интегралов по траекториям~\cite{Feynman88qedRus}. Дополнительное рассмотрение~\cite{Kisil12cRus,Kisil17a} показывает, что именно \emph{периодичность} функции \(\exp(\rmi \omega t)\) лежит в основе квантовой механики, а вовсе не некоторая <<некоммутативность квантовых наблюдаемых>>, как было постулировано Дираком в~\cite{Dirac26b}.
\item  В той же работе~\cite{deBroglie23} де~Бройль рассматривает \emph{фиктивную} пространственную волну такую, что в каждой точке пространства-{}времени её фаза совпадает с фазой внутреннего периодического процесса каждого кванта находящегося в этой точке. Это предположение \emph{стабильности} (также можно сказать \emph{согласованности} или \emph{когерентности}) позволило де~Бройлю впервые вывести правила квантования Бора, которые до этого приходилось постулировать.
\item \label{item:coherence-rus}
  Из предыдущего правила согласования фиктивной волны и внутреннего периодического процесса мы также можем заключить, что два кванта из согласованного ансамбля прибывшие в одну пространственно-{}временную точку разными траекториями должны иметь там  одну и ту же фазу. Сходным образом, два кванта расходящиеся из одной точки по разным траекториям начинают свои пути с общей для них фазой. 
\item Идеи  де~Бройля были использованы  Шрёдингером в~\cite{Schrodinger26b} для обоснования  волнового уравнения носящего его имя. Основываясь на оптико-{}механической аналогии Гамильтона~\cite{deBroglie26a,deBroglie10r}, которая связывает классическую механику с геометрической оптикой лучей\footnote{Для нас, обучавшихся по прекрасной книге Арнольда~\cite[\S\,46]{Arnold89Rus} может быть несколько неожиданным, что Шрёдингер называет вывод формализма Гамильтона из оптической аналогии <<почти забытым в наше время>>. Однако, мы не будем этому сильно удивляться, если заметим, что обоснование волнового уравнения, сделанное Шрёдингером, сейчас также отсутствует в учебниках---оно или сразу постулируется, или обосновывается такими же  постулируемыми <<правилами квантования>>.
}, Шрёдингер предположил, что квантовая механика аналогична более физической волновой оптике. Обсуждение этой аналогии в~\cite{Schrodinger26b} подсказывает, что частота внутреннего периодического процесса кванта является функцией времени и её производная по времени пропорциональна Лагранжиану частицы.
\item Если частота колебаний пропорциональна Лагранжиану \(L\), то изменение фазы вдоль траектории \(\gamma\) будет пропорционально функции действия  \(S=\int_\gamma L\,\rmd t\). В явном виде это уже высказано Дираком в работе~\cite{Dirac33a} и впоследствии стало основой метода интегралов по траекториям Фейнмана~\cite{Feynman88qedRus,FeynHibbs65}. Хорошо известно, что первоначально подход Фейнмана был холодно встречен сторонниками копенгагенской интерпретации   из-{}за использования запрещённого понятия <<траектория>>.  Неприятия было ослаблено только в рамках соглашения, что отдельные траектории являются лишь вычислительными фикциями и <<в действительности>> каждый квант путешествует по ним всем одновременно. Конечно, в академическом издании~\cite{FeynHibbs65} не могло быть и речи, что фаза \(S/\hbar\)  может быть приписана какому-{}либо внутреннему  периодическому процессу отдельного кванта проходящему конкретную траекторию.  Тем не менее, именно такой образ создают <<вымышленные часы, запускаемые на время, пока летит фотон>> в популярном изложении~\cite{Feynman88qedRus}, которое было не так сильно сковано официальными рамками.
\end{enumerate}

Введённое в п.\ref{item:coherence-rus} условие согласования квантового ансамбля является достаточно жёстким. В реальной ситуации точное совпадение фаз может быть ослаблено до всего лишь их \emph{достаточной} близости. Соответственно,  можно ввести меру рассогласования ансамбля суммируя все различия фаз в точках пересечения траекторий. Также возможно рассмотреть взаимодействие квантов при столкновениях, которое будет приводить к увеличению согласованности всего ансамбля  подобно движению физического тела к термодинамическому равновесию.

Таким образом мы приходим к
\begin{definitionR}
  \emph{Квантовый ансамбль} состоит из траекторий проходящих чрез некоторую область пространства-{}времени. Каждой точке траектории соответствует \emph{фаза} пропорциональная функции действия. В достаточно \emph{согласованном ансамбле} фазы любых двух траекторий встречающихся в какой-{}либо точке пространства-{}времени должны быть достаточно близки.    
\end{definitionR}
Наше использование квантовых траекторий вместо локализации частицы в пространстве-времени  сходно с рассмотрением ансамблей \emph{квантовых процессов} в книге Никольского~\cite[\S3.3, \S12]{NikolskyKV41}. 

Хорошо известно, что интегрирование по траекториям позволяет вычислить волновую функцию квантового объекта совпадающую с результатами квантования по Шрёдингеру или Гейзенбергу--Дираку~\cite[\S\,3.4]{FeynHibbs65}. Более того, подавляющее число траекторий взаимно уничтожают свои вклады в волновую функцию потому, что они прибывают в одну и ту же точку пространства-{}времени с практически противоположными фазами. Таким образом, та же самая волновая функция получается, если такие взаимно компенсирующие пути будут совместно исключены из рассмотрения. Оставшиеся пути в каждой точке будут иметь фазы близкие к фазе результирующей волновой функции и, поэтому, ансамбль будет достаточно согласован. Это рассуждение показывает, что \emph{для любой волновой функции существует достаточно согласованный ансамбль, который  производит эту волновую функцию суммированием по Фейнману}.

Противники ансамблевой интерпретации часто приводят пример \emph{сильно разреженных} ансамблей. Например, для интерференции электрона на двух щелях обсуждают прохождение электронов <<по одному>>: источник не испускает очередной электрон до тех пор пока не будет поглощён предыдущий~\cite{AspectGrangier87a}. Действительно, в этом случае отсутствуют пересечения путей квантов между собой и нет необходимости (да и возможности) согласовывать их напрямую.  Однако, в таком опыте всегда присутствует экспериментальная установка, в частности, её часть вызывающая интерференцию. Её мировая линия пересекает траектории всех квантов и, потому, должна быть частью условия согласования ансамбля. Концептуальная возможность возникновения волновых эффектов для разреженных ансамблей помещённых в такой экспериментальный \emph{контекст} была показана в работах~\cite{Kisil01c,KhrenVol01,KisilHodgson21a}.


\section{Обсуждение и приложения}
\label{sec:--dicussion}

Корпускулярно-волновой дуализм материи остаётся камнем преткновения различных  подходов к квантовой механике. Среди предлагаемых решений можно найти достаточно радикальные,  например, вообще отказать корпускулярным свойствам света в физическом существовании и объявить <<фотон>> неудачной фикцией на вроде <<эфира>> XIX века~\cite{Lamb95a}. 

В этой заметке мы предложили наполнить ансамблевую интерпретацию волновой функции некоторыми изначальными догадками де~Бройля и усилить их более поздними методами из фейнмановского интеграла по путям. Перечислим некоторые непосредственные следствия предположенного подхода:
\begin{enumerate}
\item Наблюдение волновых явлений у квантовых (дискретных) объектов всегда носит статистических характер и, потому, оно есть совокупное свойство согласованных ансамблей. Волновые свойства никак не могут быть проявлены одной частицей или группой совсем никак не согласованных частиц. Сходным образом, скорость одной молекулы газа не содержит информации о тепловом распределении в рассматриваемом объёме, хотя такая молекула и вносит свой вклад в общую картину.  Эта аналогия может оказаться не такой уж и поверхностной если взглянуть на уравнение Шрёдингера как на уравнения теплопроводности с комплексным временем. 
\item Волновая функция соответствует согласованному ансамблю квантовых объектов, а именно её амплитуда определяется плотностью пространство-{}временной локализации объектов. Фаза волновой функции совпадает с фазой кванта в данной точке расширенного конфигурационного пространства и, по сути, совпадает с <<фиктивной волной>> де Бройля из~\cite{deBroglie23}.
\item Акт наблюдения (измерения) одного квантового объекта выявляет его отдельные физические характеристики, например, пространственно-{}временную локализацию. Это может привести к незначительному возмущению в ансамбле, но никакой полной редукции волновой функции (соответствующей всему ансамблю) при единичном измерении не происходит~\cite[\S14]{Blokhintsev83a}.
\item Из интерпретации исчезают такие парадоксы, как мгновенная передача информации в эксперименте Эйнштейна--Подольского--Розена~\cite[p.364]{Mandelshtam72} и дилемма  <<жив ли  кот Шрёдингера или он скорее мёртв>>?
\item Физический смысл имеют фазы (например, как соответствующие каким-{}то внутренним периодическим процессам) отдельных квантов продвигающихся своим путём. Напротив, волновая функция является всего лишь интегральным показателем всех таких фаз в согласованном ансамбле. Поэтому, нет нужды вводить некий физический носитель вроде поля, который представляет волновую функцию, см. упомянутую выше аналогию с тепловым распределением в физическом теле.
\item  Так же нет необходимости в каких-{}либо дополнительных <<пилотных волнах>> чтобы сделать обратный переход от волновой функции к индивидуальной траектории отдельного кванта. Такие конструкции <<двойных решений>>~\cite{deBroglie70a
  } потребовались де~Бройлю после того, как он был вынужден отказаться от своих первоначальных идей под давлением сторонников копенгагенской
интерпретации~\cite[гл.\,6]{Lochak99a}, \cite[гл.~I.6]{deBroglie10r}.
\end{enumerate}

Легко заметить, что все предложенные рассуждения базируются на традиционных методах квантовой механики: волновом уравнении Шрёдингера и интеграла по траекториям Фейнмана подсвеченных некоторыми ранними предложениями де~Бройля. Это избавляет нас от необходимости делать здесь тщательную проверку на соответствие нашей интерпретации устоявшемуся математическому формализму волновой механики.

Заметим также, что в главном наше рассмотрение перекликается с подходом А.~Хренникова к ансамблевой интерпретации, см.~\cite{Khrennikov12a}: вначале ввести некоторую базовую <<до-{}квантовую>> теорию обладающую реализмом и детерминизмом, а потом получить  вероятностное описания процесса путём редукции к  волновой функции. Однако, мы (в отличие от~\cite{Khrennikov12a})  не вводим какой-{}либо новой до-{}квантовой теории, а вместо этого пытаемся пересобрать конструкцию из элементов, уже давно присутствующих в теме.  

Несмотря на то, что мы обсуждаем здесь <<всего лишь>> вопросы интерпретации, это имеет непосредственное практическое значение. Сейчас активно исследуются возможности квантового шифрования или квантовых вычислений, которые существенно базируются на предположении, что единичный квантовый объект (кубит) способен вместить в себя всю полноту волновой функции. На эти исследования, учитывая их анонсированные приложения в финансовых и военных сферах, были уже (вы)брошены огромные ресурсы. Поэтому, отсутствие каких-{}либо осязаемых результатов воплощенных в практических устройствах уже можно засчитать \emph{экспериментальной проверкой} различных теоретических   возражений высказанных ранее~\cite{Kisil09b,Dyakonov12a,Dyakonov18a,KisilHodgson21a}. В частности, такие технологии в принципе не будут работоспособны, если окажется, что волновая функция действительно является только совокупным свойством целого согласованного ансамбля.  Поэтому, отсутствие функционального квантового компьютера косвенно подтверждает ансамблевую интерпретацию подобно тому, как закон сохранения энергии выводится из многочисленных  (зачастую остроумных, но) бесплодных попыток построить вечный двигатель. 

Заключительное наблюдение, которое мы можем сделать из истории с идеям де~Бройля, расширяет список причин научного недостроя,  указанных в эпиграфе к этой заметке: некоторые элементы не находят подобающего им места ещё и по вполне субъективным обстоятельствам.

\renewcommand{\acknowledgmentsname}{Благодарности}
\begin{acknowledgments}
  Автор выражает благодарность анонимному рецензенту за полезные замечания, направленные на улучшение текста статьи. Проф.~Печенкин указал нам несколько важных источников по теме. 
\end{acknowledgments}
\small

\providecommand{\noopsort}[1]{} \providecommand{\printfirst}[2]{#1}
  \providecommand{\singleletter}[1]{#1} \providecommand{\switchargs}[2]{#2#1}
  \providecommand{\irm}{\textup{I}} \providecommand{\iirm}{\textup{II}}
  \providecommand{\vrm}{\textup{V}} \providecommand{\cprime}{'}
  \providecommand{\eprint}[2]{\texttt{#2}}
  \providecommand{\myeprint}[2]{\href{#1}{\texttt{#2}}}
  \providecommand{\arXiv}[1]{\myeprint{http://arXiv.org/abs/#1}{arXiv:#1}}
  \providecommand{\doi}[1]{\href{http://dx.doi.org/#1}{doi:
  #1}}\providecommand{\CPP}{\texttt{C++}}
  \providecommand{\NoWEB}{\texttt{noweb}}
  \providecommand{\MetaPost}{\texttt{Meta}\-\texttt{Post}}
  \providecommand{\GiNaC}{\textsf{GiNaC}}
  \providecommand{\pyGiNaC}{\textsf{pyGiNaC}}
  \providecommand{\Asymptote}{\texttt{Asymptote}}
  \providecommand{\noopsort}[1]{} \providecommand{\printfirst}[2]{#1}
  \providecommand{\singleletter}[1]{#1} \providecommand{\switchargs}[2]{#2#1}

\end{document}